\documentclass{article} 
\usepackage[final]{colm2026_conference}

\usepackage{microtype}
\usepackage{hyperref}
\usepackage{url}
\usepackage{booktabs}
\usepackage{amsmath}
\usepackage{graphicx}
\usepackage{lineno}

\definecolor{darkblue}{rgb}{0, 0, 0.5}
\hypersetup{colorlinks=true, citecolor=darkblue, linkcolor=darkblue, urlcolor=darkblue}

\newcommand{\nModels}{36}
\newcommand{\nAnalyzed}{34}
\newcommand{\nPersonas}{300}
\newcommand{\nDays}{6}

\newcommand{\nDevelopers}{nine}          
\newcommand{\paramLo}{1B}                
\newcommand{\paramHi}{671B}             
\newcommand{\nomoDir}{84.3\%}            
\newcommand{\nomoRangeLo}{75\%}
\newcommand{\nomoRangeHi}{100\%}
\newcommand{\betweenConv}{55.5\%}        
\newcommand{\withinConv}{72.6\%}         
\newcommand{\bestVarFid}{0.043}          
\newcommand{\bestVarFidModel}{Gemma-2-27B}
\newcommand{\gptFiveVarFid}{0.061}       
\newcommand{\gemmaFourVarFid}{0.427}     
\newcommand{\gemmaFourConv}{74.6\%}      
\newcommand{\nLevels}{6}                 
\newcommand{\lTwoPass}{26}               
\newcommand{\lSevenPass}{34}             
\newcommand{\maxLevelsPassed}{3}         

\newcommand{\cultMinInvar}{0.014}
\newcommand{\cultMaxInvar}{0.038}
\newcommand{\cultConv}{75\%}
\newcommand{\relYMean}{0.97}             
\newcommand{\withinRbaseline}{0.023}     
\newcommand{\nObsFull}{1{,}800}          
\newcommand{\dirFull}{71\%}              
\newcommand{\rhoFull}{0.38}              
\newcommand{\dirDeanchor}{77\%}          
\newcommand{\rhoDeanchor}{0.46}          
\newcommand{\withinRtempEight}{0.059}    
\newcommand{\withinRtempZero}{0.060}     
\newcommand{\miniRho}{0.45}              
\newcommand{\nanoRho}{0.48}              
\newcommand{\nanoDir}{75\%}              
\newcommand{\exhaustICC}{0.34}\newcommand{\exhaustWithinR}{0.058}\newcommand{\exhaustDir}{100\%}
\newcommand{\fatigueICC}{0.40}\newcommand{\fatigueWithinR}{0.075}\newcommand{\fatigueDir}{100\%}\newcommand{\fatigueRho}{0.95}
\newcommand{\flowICC}{0.48}\newcommand{\flowWithinR}{0.069}
\newcommand{\nIccPairs}{8}
\newcommand{\miniBaseComp}{0.693}\newcommand{\miniTunedComp}{0.702}
\newcommand{\nanoBaseComp}{0.660}\newcommand{\nanoTunedComp}{0.673}
\newcommand{\nTuneTrain}{30}\newcommand{\nTuneTest}{30}\newcommand{\nVerifiedL}{5}

\makeatletter
\def\section{\@startsection{section}{1}{\z@}{-1.0ex plus -0.4ex minus -.2ex}{0.6ex plus 0.2ex}{\large\bf\raggedright}}
\def\subsection{\@startsection{subsection}{2}{\z@}{-0.9ex plus -0.3ex minus -.2ex}{0.4ex plus .1ex}{\normalsize\bf\raggedright}}
\makeatother

\title{Benchmarking Psychological Dynamics in Generative Agents}

\author{%
\end{tabular}\hfill
\begin{tabular}[t]{c}\bf
Sumer S. Vaid\thanks{Corresponding author: \texttt{svaid@hbs.edu}}\\
AI Institute\\
Harvard Business School
\end{tabular}\hspace{0.06\textwidth}%
\begin{tabular}[t]{c}\bf
Ashley V. Whillans\\
Negotiation, Organizations \& Markets Unit\\
Harvard Business School
\end{tabular}\hfill
\begin{tabular}[t]{l}}

\begin{document}

\ifcolmsubmission
\linenumbers
\fi

\maketitle

\lhead{Published at the AI Measurement Science Workshop at COLM 2026}

\begin{abstract}
Large language models (LLMs) are increasingly deployed to simulate human behavior, acting as computational replicas of human subjects. Yet the lived psychological experience of humans is difficult to benchmark,
particularly as it unfolds over time. We introduce a psychometric benchmark for \emph{computational replicas}: personas that carry a
fixed identity through an evolving sequence of events. Built entirely from published norms and meta-analytic effects, the benchmark scores two dimensions of
psychological realism. The first, internal validity, quantifies whether generated
trajectories reproduce the internal structure of repeated human measurement: distributions, the
between- versus within-person variance partition, temporal dependence, and range. The second, external
validity, quantifies whether replicas recover established trait, state, and indicator relations. Across \nModels{} open-weight
and proprietary LLMs from \nDevelopers{} developers (\paramLo{}--\paramHi{} parameters), most recover
the \emph{direction} of established relations (\nomoDir{} mean agreement) and the variance partition
(\lTwoPass{} of \nAnalyzed{}), yet the \emph{within-person correlation} and distributional
structure elude recovery. Internal validity is independent of scale and capability: a mid-size open
LLM (Gemma-3-27B) strikes the best trade-off between the two dimensions. The benchmark is a precondition
for using computational replicas in causal inference across domains (e.g., marketing, healthcare), and identifies within-person grounding as the central challenge ahead.
\end{abstract}

\section{Introduction}

One of the most promising applications of AI is the ability to simulate how humans think, feel and behave \citep{argyle2023, grossmann2023, serapiogarcia2023}. Large language models are increasingly deployed to model cognitive processes, economic
behaviors, and social interactions, effectively acting as computational replicas
for human subjects \citep{horton2023, binz2025, park2023}. The scientific standards
governing the evaluation and quantifying the realism of computational replicas have received surprisingly limited empirical attention~\citep{ye2025psychometrics}.  Despite the proliferation of benchmarks that peg the performance of AI on specific tasks, it remains challenging to quantify the relative success of different technical approaches in recreating the psychological richness of humans' lived experiences \citep{santurkar2023, atari2023}.

Historically, the evaluation of personas has relied on static,
cross-sectional measurements \citep{simbench2026}. LLMs are prompted to
adopt an identity and respond to an isolated query, yielding a single-point
estimate of behavior \citep{park2024}. Yet human psychology is intrinsically dynamic
\citep{nesselroade1991},  continuously
adapting to accumulating contextual feedback. Relying on isolated snapshots
limits our capacity to build valid computational models of complex behavior.
For LLMs to function as reliable simulators of human psychology, they
must demonstrate realism in how their patterns of thinking, feeling and behaving change over time  \citep{stablepersonas2026}. They
must possess the capacity to mimic how humans process sequences of interlaced events, integrate various contextual changes, and reflect the cumulative impact of time on psychological states.

We introduce a new paradigm to benchmark the psychological dynamics of computational replicas. In this paradigm, computational replicas are endowed with a
stable identity and exposed to an evolving sequence of environmental
events. By measuring the persona's psychological state at each
timestep, we capture a continuous trajectory rather than a static
snapshot. As AI-based systems scale from isolated experimental replications
\citep{argyle2023, aher2023, horton2023} to person-specific
computational replicas \citep{park2024}, isolating computational cognition from
statistical artifact becomes a paramount scientific necessity. A
scientifically viable persona must demonstrate both core psychological stability
and adaptive, contextually grounded dynamics. This verification lets the field ground a science of
synthetic behavior in valid simulations of human mechanics rather than the biases of the underlying
LLM.

We introduce a comprehensive psychometric
benchmark designed to evaluate longitudinal psychological state trajectories
within synthetic populations. We define psychological realism through two
distinct dimensions: whether generated trajectories reproduce the
internal dynamics of repeated human measurements, and whether they
recover established psychological relations documented in the past literature. We
execute this benchmark on a simulated cohort of \nPersonas{}
population-grounded synthetic employees observed across \nDays{} continuous
workdays. By evaluating \nModels{} open-weight and proprietary LLMs---drawn from
\nDevelopers{} developers and spanning nearly three orders of magnitude in scale
(\paramLo{}--\paramHi{} parameters)---tasked with generating daily measures of work
engagement, psychological safety, creativity, and team trust, we establish a highly
auditable baseline. This framework provides the necessary
computational infrastructure to compare how accurately temporal computational replicas
and other generative simulations reproduce valid, context-sensitive
psychological states over time. Validating these individual dynamics is a
strictly necessary precursor to robust causal inference, a core use case for
these emerging technologies. Valid counterfactual simulation requires
an accurate structural representation of the underlying temporal
data-generating process \citep{pearl2009, hernan2020}. With validated within-person dynamics, a persona can address causal questions about change over time, such as how a shift in workload would move employee engagement over a week \citep{vaid2026testbed}.

\section{Related Work}

\subsection{The Limits of Cross-Sectional Evaluation}
The computational modeling of human behavior initially demonstrated remarkable
success in replicating static population distributions. Early frameworks
established that LLMs could output survey responses and replicate
behavioral experiments by matching macroscopic cross-sectional data
\citep{argyle2023, aher2023, horton2023, binz2023}. This capacity quickly expanded into generating
synthetic data for human-computer interaction research \citep{hamalainen2023},
where LLMs effectively substituted for human crowdsource workers on discrete
tasks \citep{dillion2023}. However, recent methodological critiques highlight that evaluating these
LLMs frequently defaults to single-trial, response-level elicitation
\citep{cummins2025}. This reliance on isolated snapshots introduces unconstrained
analytic flexibility, masking whether a LLM possesses stability or
merely outputs context-dependent artifacts tailored to a specific prompt string.

This limitation is particularly evident in recent comprehensive cross-sectional
benchmarks. Frameworks such as SimBench \citep{simbench2026} and OdysSim \citep{odyssim2026} provide
extensive infrastructure for evaluating how well LLMs replicate human behavior across diverse
populations and tasks. Yet, investigations into
survey-response consistency \citep{dominguezolmedo2024} and one-shot persona
expression \citep{jiang2024} demonstrate that matching a cross-sectional
distribution leaves temporal continuity an open question. These static paradigms
successfully validate localized demographic matching but conflate a LLM's
capacity to output a probable text string with the possession of a stable,
underlying behavioral architecture \citep{validity2025, vendrow2025}. Static framing also stands apart
from interactive measurement, where adaptive psychometric methods select items by their information
content and estimate a LLM's trait position as responses accumulate \citep{atlas2026, metabench2025}.

\subsection{Alignment Compression and Statistical Priors}
Transitioning from static generation to longitudinal simulation exposes
trade-offs in the underlying alignment of these architectures. As
LLMs undergo instruction tuning and safety alignment, they suffer a
measurable reduction in conceptual and behavioral diversity \citep{murthy2025, kirk2024, homogenization2026}.
We define this compression as a measurable loss of behavioral diversity:
alignment forces the distribution of generated responses toward a normative
consensus. This phenomenon mirrors the alignment-simulation tradeoff identified
in cross-sectional benchmarks, where instruction tuning improves performance on
consensus-driven questions but actively degrades the LLM's ability to
simulate diverse edge cases \citep{simbench2026}.

This compression is a consistent bias rather than random
variance. The resulting loss of variance is observable in LLMs exhibiting
pronounced sycophancy \citep{sharma2024} and artificially elevated baseline
rationality \citep{liu2025}, and in the documented compression of opinion and
trait distributions toward the center of the training corpus
\citep{santurkar2023, bisbee2024}. Because alignment naturally reduces
cognitive diversity, the generated personas cluster at
the center of the training distribution. When applied to complex environments
like organizational simulation, the erasure of individual variance eliminates
the diversity necessary to model realistic human dynamics
\citep{wang2025identity, kolluri2025}. Recent work reframes this failure as a
problem of support coverage rather than density matching, evolving persona
generators that span the full landscape of human traits and recover the variance
that standard alignment removes \citep{paglieri2026}. Rigorous longitudinal measurement is what lets
researchers distinguish valid simulated behavior from the generative compression inherent to the base
LLM \citep{gao2025caution}. A related validity threat arises under strategic optimization: a persona that
infers it is under evaluation may skew its self-reports toward desirable poles \citep{needham2025,
salecha2024} or deliberately underperform when an objective rewards it \citep{vanderweij2024, meinke2024,
greenblatt2024}.

\subsection{Architectural Bottlenecks Over Time}
Preliminary attempts to sustain personas reveal architectural
constraints in processing long-term state changes. While LLMs
can express specific personality traits in isolated evaluations
\citep{jiang2024, serapiogarcia2023, shanahan2023}, simulating these traits over
extended interactions exposes systematic drift
\citep{stablepersonas2026, coherent2025}. Formal analyses of agent-based social simulations
indicate that personas reliably decay toward an average, homogenous identity as
context lengths increase \citep{taillandier2025, spasm2026}. This homogenization suggests
that current LLMs struggle to maintain specialized,
individual-level variance over extended temporal horizons.

Mechanistic investigations attribute this decay directly to inherent
working-memory limits within transformer architectures \citep{gong2024}.
Although constrained repeated-game scenarios show initial promise in following
path-dependent variables like trust \citep{xie2024}, discrete prompt
engineering can stabilize specific value alignments \citep{rozen2025}, and
multi-turn reinforcement learning can reduce persona inconsistency
\citep{abdulhai2025}, these
interventions act as localized patches rather than structural solutions. More
recent training approaches target internal state alignment rather than surface
imitation \citep{humanlm2026}, yet they optimize for response accuracy and stop short of
certifying coherence across time. The
broader absence of temporal stability complicates the deployment of complex
synthetic populations. Moving from heuristic prompt adjustments to formal
psychometric validation \citep{ye2025psychometrics} is a mandatory step to ensure personas maintain
coherent identity trajectories.

\subsection{The Necessity of Longitudinal Measurement}
Despite these limitations, the field is rapidly advancing toward
multi-agent paradigms \citep{park2023, agentsociety2025}. Theoretical frameworks increasingly position
multi-agent systems as a new paradigm for social science
\citep{haase2025, anthis2025}. Furthermore, researchers hypothesize that single
LLMs can act as universal substrates to predict and capture
cross-domain human cognition \citep{binz2025}. These theoretical LLMs presume
that the underlying personas possess the capacity to integrate
experience, update their internal states, and exhibit path-dependent
development over time. They assume the existence of a stable temporal
architecture that still awaits psychometric verification.

In practice, applied systems are already being deployed in high-stakes
longitudinal contexts. Computational replicas are increasingly utilized in
prolonged educational simulations \citep{mannekote2025} and continuous
psychological counseling frameworks \citep{hu2025theramind}. These deployed
systems operate under the unvalidated assumption of continuous temporal
coherence, and they contend with an instrument that itself drifts: deployed LLMs change across
silent updates \citep{chen2024drift} and the preferences they encode shift over time \citep{son2025},
so a longitudinal standard must separate intended development from artifactual movement of the
underlying LLM. A nascent literature on validating simulations as behavioral evidence
has begun to formalize what such validation requires, distinguishing several
strategies and their limits \citep{hullman2026, hewitt2024}, while related
analyses caution that distributional realism can coexist with incorrect causal
effects \citep{cui2025, causal2026}. Our framework addresses this precise methodological void. By
establishing a rigorous standard for measuring longitudinal coherence, we
provide the computational precondition required to advance beyond static
evaluations. This auditing infrastructure secures the architectural foundation
necessary to validate temporal computational replicas \citep{park2024, toubia2025, socialdt2026} and ensure they function as
reliable engines for longitudinal simulation.

\section{Creating the Benchmark}

\subsection{Population Initialization}
We initialize synthetic populations from published norms rather than LLM guesswork, so every seed
parameter traces to a human dataset rather than the generation LLM's assumptions \citep{gao2025caution}.
Each persona's Big Five traits are drawn from a multivariate normal grounded in population norms and
meta-analytic intercorrelations \citep{soto2011, vanderlinden2010}, and state baselines are linear
functions of those traits using established coefficients \citep{christian2011, frazier2017, feist1998,
colquitt2007}. We decompose each state's total variance into stable between-person and dynamic
within-person components by scaling its standard deviation with published intraclass correlations,
following the experience-sampling tradition \citep{nesselroade1991, podsakoff2019}. The
benchmark targets four workplace states: work engagement, the vigor and absorption a person brings to
work \citep{schaufeli2002}; psychological safety, a shared confidence that the team tolerates
interpersonal risk \citep{edmondson1999}; workplace creativity, the generation of novel and useful ideas
\citep{amabile1988}; and team trust, a willingness to be vulnerable to a team's actions
\citep{mayer1995}. A rich tradition in personality and organizational psychology separates such states
from traits: traits are stable between-person dispositions, while states fluctuate within the individual
around a personal baseline and aggregate over time into the trait
\citep{fleeson2001, molenaar2004, curranbauer2011}. The benchmark measures whether generative digital
twins reproduce realistic within-person fluctuation in these psychological states under salient daily context.
Daily contextual indicators are sampled forward as inputs while the target state columns
stay unpopulated, so any recovered indicator-state relation is emergent rather than seeded
\citep{gao2025caution}. We initialize \nPersonas{} personas over \nDays{} simulated days; the full
seeding construction appears in Appendix~\ref{app:seeding}.

\subsection{Persona Generation}
We employ a shared-event architecture to strictly control environmental
variance \citep{simbench2026}. For each seeded persona, we generate an identity portrait
and a continuous sequence of daily narratives conditioned on the evolving
contextual inputs. This generation occurs exactly once using a fixed event
LLM. During the evaluation phase, every benchmarked LLM processes these
identical narratives under a uniform prompt structure.

Each LLM is tasked with quantifying the four psychological states in the workplace at the
conclusion of each daily timestep, outputting values on their native
psychometric scales. Because all LLMs rate identical narratives, differences in ratings reflect the capability of the LLM rather than the events. For every evaluated LLM, this procedure
yields a complete behavioral trajectory spanning \nDays{} discrete state
vectors across the entire \nPersonas{} persona cohort.

\subsection{Two Dimensions of Psychological Realism}
The benchmark scores a synthetic population on two dimensions, following the construct-validity
distinction between a measure's internal structure and its relations to other measures
\citep{cronbachmeehl1955, campbellfiske1959}.

\emph{Internal validity} compares the synthetic corpus to published human norms on the statistical
properties of experience-sampling data, using two one-sided equivalence tests with persona-level
resampling \citep{lakens2017}. The criteria are the mean and spread of each state, the between- and
within-person variance partition, within-person variability, cross-state correlations, the within-person
autoregressive coefficient, latent dimensionality, and scale-range use; Appendix~\ref{app:levels} defines
each. Together these properties describe how a psychological state behaves under repeated measurement,
and a faithful computational replica must match them all. We report two criteria in adjusted form because
the design limits their identification: a one-factor model over four indicators is under-identified
\citep{kline2016}, so we report multilevel eigenstructure in its place, and over \nDays{} days the
autoregressive coefficient is biased by the short panel \citep{nickell1981, hamaker2015}, so we treat it
as directional.

\emph{External validity} examines whether the relationships in the synthetic data match those established in
human research, at two levels. Between persons, we test whether a persona's traits predict
their average states (their \emph{trait--state} relations), as meta-analyses report; for example, more
conscientious personas should be more engaged. Within persons, we test whether a persona's daily context
predicts the state they report that day (their \emph{indicator--state} relations), by correlating each
day's indicators with each day's states after subtracting the persona's own average
\citep{enderstofighi2007, curranbauer2011}. A known-groups test adds a final check: more neurotic
personas should report lower engagement on average and larger day-to-day swings \citep{fleeson2001}.

\section{Experimental Setup}

We deploy this benchmark across a broad roster of \nModels{} LLMs drawn from
\nDevelopers{} developers and spanning \paramLo{}--\paramHi{} parameters, nearly
three orders of magnitude in scale. This cohort spans dense parameter ladders,
state-of-the-art mixture-of-experts architectures, specialized reasoning engines,
and both open-weight and proprietary frontier systems. We analyze the \nAnalyzed{}
LLMs that produced complete longitudinal corpora.

As established in the shared-event protocol, identity portraits and sequential
daily narratives are generated exactly once using a fixed event model.
Every benchmarked LLM then processes this shared corpus under a strictly
uniform prompt architecture. To execute this at scale, the generation and
rating pipelines operate concurrently and are fully resumable. The
orchestration infrastructure includes integrated budgetary halts and an
automated error-rate circuit breaker to ensure systemic stability during
large-scale inference.

We quantify external nomological convergence using direction agreement, defined
as the proportion of simulated relations where the algebraic sign correctly
matches the established literature, alongside signed rank correlations across
the complete relation set. Internal validity is summarized through
rigorous two one-sided tests equivalence verdicts \citep{lakens2017} and continuous distance
metrics. The primary continuous diagnostic is the \emph{within-person variance
error}: the absolute log-ratio of the synthetic within-person standard deviation
to the target human standard deviation. In this formulation, lower absolute
values indicate a closer match to the human within-person variance structure.

\section{Results}
\label{sec:results}
\begin{figure}[t]
\centering
\includegraphics[width=0.95\linewidth]{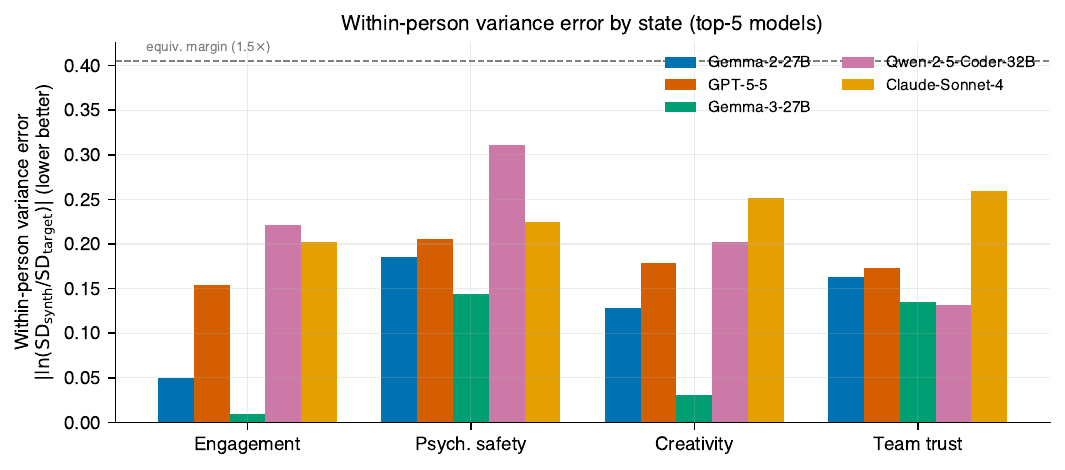}
\caption{Internal validity by psychological state for the top-five LLMs: the absolute log-ratio of
synthetic to human within-person SD, where lower is closer to the human norm. The dashed line marks
the equivalence margin ($1.5\times$). LLMs match the within-person variance of engagement and trust
more closely than psychological safety.}
\label{fig:leaderboard}
\end{figure}

\begin{figure}[t]
\centering
\includegraphics[width=0.95\linewidth]{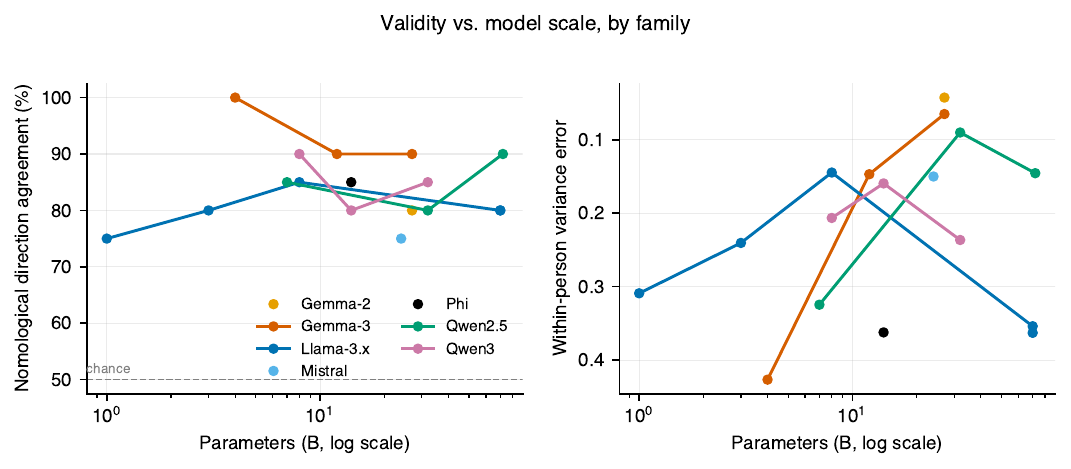}
\caption{Validity versus LLM scale, by family. External validity (left, direction agreement) is high
and roughly flat across parameter count; internal validity (right, within-person variance
error) varies sharply and is led by mid-size LLMs rather than the largest.}
\label{fig:scaling}
\end{figure}

We report the outcomes of our diagnostic framework across the \nAnalyzed{} evaluated LLMs along two
axes. \emph{Internal validity} examines whether a synthetic population reproduces the longitudinal
structure of repeated human measurement---most directly, whether its day-to-day, within-person
variability matches the human norm. \emph{External validity} examines whether the population recovers
established psychological relations in the correct direction. Figure~\ref{fig:leaderboard} reports
internal validity by psychological state for the leading LLMs; Figure~\ref{fig:scaling} reports both
axes against LLM scale.

\subsection{The Divergence of Scale and Internal Validity}
A naive observer, extrapolating from scaling laws \citep{kaplan2020, hoffmann2022} and the broad
capability gains larger models display \citep{metabench2025, binz2025}, would assume that larger and more
capable LLMs produce synthetic populations whose psychological states display high internal validity
(e.g., human-like day-to-day variability) and external validity (e.g., higher neuroticism predicting lower engagement). A larger LLM attains lower predictive loss by capturing more of what its training data contains: the
documented relations that external validity measures, and the distributional and temporal regularities
that internal validity measures. On both axes, a higher parameter count should therefore buy more
faithful trait--state relations and more realistic within-person dynamics. Our benchmark shows
otherwise. Fitting each validity metric against log parameter count within LLM families, we find that
general capability and internal validity move largely independently.

\begin{figure}[t]
\centering
\includegraphics[width=0.62\linewidth]{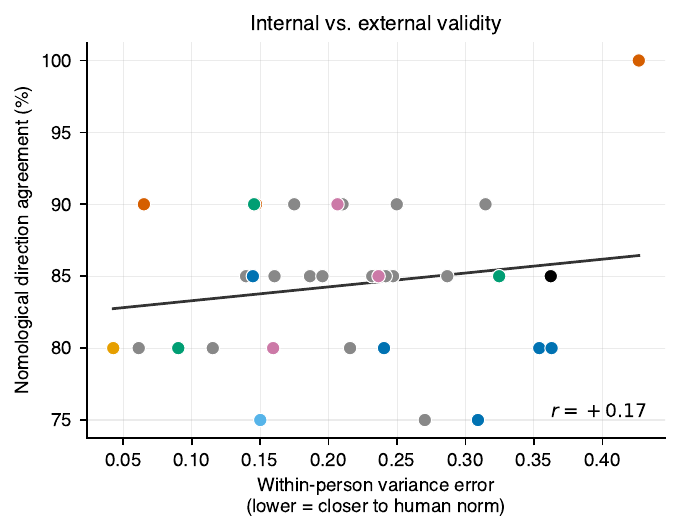}
\caption{Internal versus external validity, one point per model. The near-zero slope
($r \approx 0.17$) shows the two axes are largely independent: gains in internal
validity leave external validity unchanged.}
\label{fig:tradeoff}
\end{figure}

We observe sharp divergences in how different architectures process temporal sequences. The smallest
within-person variance error is attained by a mid-size open
LLM (\bestVarFidModel{}, \bestVarFid{}, where lower values indicate closer alignment to human
targets), narrowly ahead of the most mathematically capable frontier LLM (\gptFiveVarFid{}).
Conversely, the most compact Gemma architecture exhibits the largest variance error in the
clean roster (\gemmaFourVarFid{}) while still matching the direction of established relations at a rate
comparable to far larger LLMs (\gemmaFourConv{} pooled directional agreement). No single architecture
leads across both internal and external metrics. External validity remains remarkably constant across
the capability range, whereas internal validity fluctuates unpredictably; across the roster the two
axes are largely independent (Appendix Figure~\ref{fig:tradeoff}, $r \approx 0.17$). The internal battery
separates what LLMs reproduce from what they miss. The between- versus within-person variance partition
(L2) and the range criterion (L7) are met by most of the roster (\lTwoPass{}, \lSevenPass{} of
\nAnalyzed{} populations); distributional shape (L1), cross-state correlations (L3), the full
multilevel eigenstructure (L6) are met by few. The best population clears
\maxLevelsPassed{} of the \nLevels{} levels, so the synthetic populations reproduce the variance
architecture of repeated human measurement while stopping short of its distributional and relational
detail. Two criteria warrant added caution: the cross-state (L3) targets remain literature placeholders
rather than verified anchors, and the within-person temporal criterion (L4) is reported as directional
evidence only---at \nDays{} days the autoregressive coefficient is non-identifiable, so credible
estimation of persona \emph{dynamics} requires longer trajectories than the experience-sampling horizon
used here.

\subsection{Compression of the Within-Person Correlation}
While the synthetic populations demonstrate a robust capacity to recover the directional signs of
established psychological relations, they systematically fail to recover the within-person correlation
that these effects imply. Recovering direction is distinct from recovering that correlation. The trait--state network is
recovered well: signs align with published coefficients in a mean \nomoDir{} of cases (range
\nomoRangeLo{}--\nomoRangeHi{}), the strongest external-validity signal in the benchmark. Within
persons, daily indicators move states in literature-consistent directions in approximately
three-quarters of relations (\withinConv{}).

Yet the simulated effects are compressed toward zero. The pooled within-person
correlations hover near an absolute value of $0.05$, whereas the published human effects range from
$0.20$ to $0.70$. This compression of the within-person correlation is uniform across the entire capability spectrum,
rather than a localized failure of smaller LLMs. These LLMs
correctly identify that a specific environmental indicator should push a psychological state in a
particular direction, yet they lack the capacity to drive that state change with the
appropriate statistical force. Between persons, indicator--state convergence falls to \betweenConv{},
near chance, because the seeded daily indicators carry no trait-linked between-person variance by
construction; the trait--state network, rather than the indicator network, is therefore the population-level
external check. We probed whether this compression is an artifact of the rating protocol. Re-rating
identical narratives yields a single-rating reliability of \relYMean{}, so the reliability correction is
negligible; the within-person correlation is moreover invariant to rating temperature and to a battery
of prompt interventions that de-anchor the rater from its trait baseline. Across every condition the
within-person correlation is unmoved and sits at the sampling noise floor (\withinRbaseline{} at \nObsFull{}
person-days). The compression is therefore an inherent property of the synthetic process rather than an
instrument effect: the within-person daily signal is intrinsically faint. Its direction and rank are
nonetheless recovered (\dirFull{}--\dirDeanchor{} sign agreement, $\rho=\rhoFull{}$--\rhoDeanchor{}
against meta-analytic effects), and removing the baseline anchor sharpens this directional recovery
while leaving the within-person correlation untouched. The quantity at issue is thus the within-person \emph{structure} of
the synthetic trajectories measured against published effects; recovering the realized state movements
of specific human individuals is a distinct and harder problem that demands longitudinal ground truth
and lies beyond an all-silico evaluation.

\subsection{Cross-Cultural Invariance and Demographic Simulation}
To test the resilience of the generated trajectories across diverse populations, we
re-initialized the personas for six distinct nations using Hofstede-informed trait offsets and
re-executed the diagnostic battery. The resulting measurement invariance is exceptionally strong. The
per-culture cross-state correlational structures depart from the baseline global structure by a mean
absolute difference of \cultMinInvar{} to \cultMaxInvar{}.

This deviation sits far below conventional threshold limits for breakdown. Furthermore, the
simulated cross-group differences match the published Hofstede-anchored directions in up to \cultConv{}
of the evaluated cases, with power-distance effects on psychological safety and creativity recovered
consistently. The capability divergence recurs: the larger Gemma-3-27B barely differentiates
cultures, whereas the smaller Gemma-3-4B differentiates them strongly. These results indicate that when
personas are anchored with precise demographic priors, the LLMs keep a consistent cross-state
correlation pattern across cultures while still distinguishing one from another.

\section{Discussion}
\label{sec:discussion}

The LLMs we evaluate recover the direction of established psychological relations with high accuracy. Building upon the foundational
cross-sectional matching demonstrated in past research \citep{argyle2023, aher2023, horton2023}, our
diagnostic framework confirms that LLMs successfully internalize the established trait--state
relations we test \citep{serapiogarcia2023, binz2025, park2023}. Moreover, the evaluated cohort achieved high external validity,
recording a mean direction agreement of \nomoDir{} across all meta-analytic targets. Furthermore,
precise demographic priors kept the cross-state correlation pattern consistent across cultures while
preserving the differences between them. These findings
indicate that current LLMs correctly identify how an environmental indicator should
influence the workplace states we study \citep{grossmann2023}.

However, this qualitative success masks a critical architectural bottleneck regarding longitudinal
variance. We observe a systematic divergence between general computational capability and internal
validity. While LLMs correctly predict the direction of a psychological shift, they consistently fail
to reproduce the within-person correlation and temporal variance of these states. This systemic compression of the within-person correlation is
a consistent bias rather than random noise. As shown in past research \citep{murthy2025},
instruction tuning actively reduces cognitive diversity. This aligns with
limitations identified across related work on alignment and automated simulation \citep{santurkar2023,
atari2023, manning2024}, where the optimization gradient forces generated responses toward a normative
consensus, thereby compressing the variance required for dynamic population simulation.

Securing temporal coherence is a strictly necessary precondition for conducting robust causal inference
with synthetic computational replicas. Valid counterfactual simulation requires an accurate
structural representation of the underlying temporal data generating process, as established in
foundational work on causal inference \citep{pearl2009, hernan2020}. Applied frameworks frequently deploy
computational replicas in longitudinal contexts under the unvalidated assumption of continuous stability
\citep{park2024, stablepersonas2026}. Yet our results illustrate that distributional realism and
directional accuracy can easily coexist with incorrect causal effect sizes \citep{simbench2026}. Moving
from heuristic prompt adjustments to formal psychometric validation ensures that the science of synthetic
behavior rests on verified psychological dynamics rather than LLM artifacts.

\section{Conclusion}

Transitioning from static personas to longitudinal computational replicas exposes a gap between directional and quantitative fidelity when
LLMs rate an extended record of experience \citep{park2024, stablepersonas2026}. By framing
longitudinal coherence as a rigorous measurement problem, we establish a highly auditable standard for
validating synthetic psychological trajectories. The demonstrated success in directional alignment
proves that LLMs hold immense promise for organizational and psychological simulation
\citep{argyle2023, grossmann2023, vaid2026testbed}. However, the simultaneous divergence between LLM scale and
within-person variance recovery mandates a shift in how the field evaluates these systems. Researchers
must verify the stability of the underlying LLMs before deploying them to simulate
populations over time.

\subsubsection*{Reproducibility Statement}
The validation battery, generation pipeline, LLM scoring code, generated corpora for all \nModels{} models, result tables with provenance manifests, and random seeds are released as PsyDyBench at \url{https://github.com/sumervaid/PsyDyBench}. A single script regenerates every table, figure, and reported number from the released corpora without API access.

\bibliography{colm2026_conference}
\bibliographystyle{colm2026_conference}

\appendix

\section{The Within-Person Correlation: Why It Compresses and How to Recover It}
\label{app:magnitude}

The benchmark recovers the direction of within-person relations, yet the within-person correlation stays near zero,
and this gap is the central puzzle of this work. In this appendix we explain why a generative
rater returns a small, central deviation when the day calls for a larger one, a pattern we call the
safe-deviation tendency, and we set out the factors that enlarge it. Throughout, we separate the routes
available to an all-silico evaluation from the routes that require longitudinal human data.

\subsection{The safe-deviation tendency}
\label{app:safe}

We trace the timid response to four mechanisms that reinforce one another. First, the rating a model returns is
closer to an average than to a single draw: asked for an end-of-day state, the model reports its central
expectation over plausible responses, and an average is smoother than the responses it summarizes, so
the reported series carries less within-person variance than the day-to-day state it stands for. The
near-deterministic ratings we observe, with single-rating reliability \relYMean{}, confirm that the
rater commits to this average, which is also why the within-person correlation holds at
\withinRtempEight{} under rating temperature 0.8 and \withinRtempZero{} under temperature 0.0. Second,
alignment rewards the modal answer: instruction tuning and preference optimization favor moderate,
broadly acceptable outputs and discourage commitment to scale extremes, so the rater settles toward the
middle of each range, a pattern we make concrete in Section~\ref{app:capability}, where general
capability and within-person recovery move in opposite directions. Third, the supplied baseline acts as
a prior: when we hand the rater its own trait baseline and typical range, the model shrinks each day
toward that prior, and removing the anchor sharpens directional recovery, lifting sign agreement from
\dirFull{} to \dirDeanchor{} and rank correlation from \rhoFull{} to \rhoDeanchor{} while the within-person correlation
holds steady, which marks anchoring as one contributing cause among several. Fourth, the signal weakens
across two mappings: a day's drivers pass from seeded indicators to narrative and from narrative to
rating, and because each mapping pulls toward its own average, the link between an indicator and the
state it should move has been smoothed twice by the time we measure it. Taken together, these mechanisms
leave a rater that names the correct sign of a day's effect and renders it as a small, safe deviation,
faithful in direction and timid in magnitude.

\subsection{Capability and within-person recovery}
\label{app:capability}

We test the role of capability directly, holding the prompt and the data fixed while re-rating the
shared narratives with three models that span the gpt-4.1 family. As capability falls, the within-person
rank signal strengthens: the rank correlation against meta-analytic effects rises from \rhoFull{} for
gpt-4.1, through \miniRho{} for gpt-4.1-mini, to \nanoRho{} for gpt-4.1-nano, and sign agreement reaches
\nanoDir{} for the smallest model. The within-person correlation, by contrast, stays near the sampling floor across the
whole ladder, between \withinRbaseline{} and 0.028. We read this as evidence that the lighter-aligned
model preserves the day-to-day variation the structure requires, while the within-person correlation remains governed by
the construct's variance rather than by model capability.

\subsection{Construct variance and within-person recovery}
\label{app:icc}

The variance available for any within-person relation is set by the intraclass correlation, and because
the four headline states are trait-saturated, with ICC near 0.75, little within-person variance is left
for an indicator to move. We therefore test three constructs with a more even between- and within-person
split, re-rating the same narratives for daily exhaustion (ICC \exhaustICC{}), fatigue (ICC
\fatigueICC{}), and flow (ICC \flowICC{}). The within-person correlation rises in step with the within-person
variance share: pooled $|r|$ reaches \exhaustWithinR{} for exhaustion, \fatigueWithinR{} for fatigue, and
\flowWithinR{} for flow, roughly three times the \withinRbaseline{} of the trait-saturated set, with sign
agreement of \exhaustDir{} and \fatigueDir{} and rank correlation up to \fatigueRho{}. These estimates
rest on \nIccPairs{} anchored pairs per state and serve as a directional demonstration, yet the pattern
is clear: the limit on the within-person correlation sits in the construct rather than in the model, and where a state carries
real day-to-day variance, the synthetic population reproduces within-person structure at a size the
literature recognizes.

\subsection{In-silico achievability of the internal-validity targets}
\label{app:achieve}

We next ask whether the internal-validity targets are reachable by optimization that draws on published norms
alone, holding measured human states out of the objective entirely. We optimize the rating prompt with
MIPROv2 against a literature-only objective that combines the within-person standard deviation implied by
published ICC values, use of the full scale range, avoidance of ceiling and floor pileup, and the
cross-state coherence band on the \nVerifiedL{} verified meta-analytic pairs. We optimize on
\nTuneTrain{} training personas and report the composite on \nTuneTest{} personas held aside from
optimization. The tuned prompt raises the held-out composite from \miniBaseComp{} to \miniTunedComp{} for
gpt-4.1-mini and from \nanoBaseComp{} to \nanoTunedComp{} for gpt-4.1-nano. We read the small and
consistent gains two ways: the uniform off-the-shelf prompt already sits close to the in-silico ceiling
for these targets, and optimization against published norms still produces a measurable
improvement on held-out data. The instruction the optimizer selects encodes the same anti-anchoring move
we identify in Section~\ref{app:safe}, directing the rater to let a strong day move each rating
toward the ends of its scale. These results establish reachability on a fully all-silico footing,
separate from the question of predicting realized individual trajectories.

\subsection{A principled bound and the routes that remain}
\label{app:bound}

We close with two readings. First, the ceiling on the within-person correlation is structural: when between-person variance
dominates the partition, the recoverable within-person correlation is bounded, and a weak within-person correlation on
trait-saturated states is the expected outcome. Second, the routes that raise the within-person correlation in an all-silico
setting act on variance and structure, and they include a lighter-aligned rater
(Section~\ref{app:capability}), constructs with a larger within-person share (Section~\ref{app:icc}),
de-anchored rating, and narratives that encode the day's drivers more vividly. Routes that condition on
an individual's measured history, namely personalized exemplars, neighbor retrieval, and fitted
statistical predictors, raise pooled coefficients while their within-person component stays near chance,
and because they depend on measured human states they belong to the companion problem of predicting
realized individual trajectories. We therefore set the honest target for an all-silico benchmark as
matching within-person variance, range, and direction to published effects, and treating the within-person correlation as a
value the construct's variance bounds.

\section{Seeding construction}
\label{app:seeding}
We initialize each persona by drawing five trait scores from a multivariate normal
$\mathcal{N}(\mu,\Sigma)$, where $\mu$ holds the published Big Five means and $\Sigma$ assembles
published standard deviations and the meta-analytic intercorrelation matrix \citep{soto2011,
vanderlinden2010}; when the assembled $\Sigma$ departs from positive definiteness, we project it to the
nearest positive-definite matrix before sampling. We then set each state's person-level baseline as a
linear function of the standardized traits, $b=\beta^{\top}z$, with $\beta$ extracted from the cited
meta-analyses \citep{christian2011, frazier2017, feist1998, colquitt2007}, and we split total state
variance with the published intraclass correlation, so that the between-person standard deviation is
$\sigma_{b}=\sigma\sqrt{\mathrm{ICC}}$ and the within-person standard deviation is
$\sigma_{w}=\sigma\sqrt{1-\mathrm{ICC}}$. Baselines sit on each state's native scale, and we leave the
target state columns unpopulated so that any recovered indicator-state relation emerges from generation.

\section{Internal validity criteria}
\label{app:levels}
We score six internal criteria, each as two one-sided equivalence tests \citep{lakens2017} against
literature targets with persona-cluster bootstrap intervals.
\textbf{L1 Distributions:} state means and spreads match published norms (Hedges' $g$ equivalence).
\textbf{L2 Variance partition:} the within-person standard deviation and the between- versus
within-person ICC split match literature targets (log-ratio equivalence).
\textbf{L3 Cross-state structure:} between- and within-person cross-state correlations match
meta-analytic pair targets (a $|\Delta r|$ margin), scored on the sourced pairs.
\textbf{L4 Temporal dependence:} the within-person AR(1) coefficient matches its target band, reported as
directional evidence given short-panel non-identifiability at six days.
\textbf{L6 Latent dimensionality:} multilevel eigenstructure (between- and within-person leading
components) stands in for a four-indicator factor model, which is under-identified.
\textbf{L7 Range utilization:} ceiling and floor pileup stays within the human range.
The numbering omits L5, an affect-specific check outside the present scope.

\section{Rating prompt}
\label{app:prompt}
Every benchmarked LLM receives the identical prompt below, assembled from the persona's identity
portrait, the day's narrative, and a fixed rating instruction. The system message places the model
inside the persona and the instruction asks for a first-person end-of-day self-report; the models do not
annotate a third party's text. No part of the prompt is tuned per model.

\begin{quote}\small\ttfamily
SYSTEM: You are the person described. Respond in first person, honestly.\\[4pt]
USER:\\
=== WHO YOU ARE ===\\
You are \{first\} \{last\}.\\
\{psychological portrait\}\\
=== YOUR REACTIVITY PATTERNS ===\\
\{reactivity patterns\}\\
=== TODAY (Day \{d\}, \{weekday\}) --- DAY TYPE: \{type\} ===\\
\{narrative\}\\
=== SCALE GUIDANCE ===\\
\{per-state scale, this persona's baseline and typical range\}\\[4pt]
=== RATE YOUR END-OF-DAY STATE ===\\
You have just lived this day. It is 5:30pm. Report your ACTUAL internal state right now
- not how you should feel, how you do feel. Honor the texture of the day.\\[4pt]
OUTPUT EXACTLY this block and nothing else (one number each, within the stated scale):\\
STATES:\\
- Work Engagement: [1-7] - [one sentence why]\\
- Psychological Safety: [1-4] - [one sentence why]\\
- Work Creativity: [1-6] - [one sentence why]\\
- Team Trust: [1-5] - [one sentence why]
\end{quote}

\section{Model leaderboard}
\label{app:leaderboard-sec}
Table~\ref{app:leaderboard} ranks every evaluated model on both criteria at once. We order by internal
validity, computing the within-person variance error against the human norm as the sort key, and we place
the external-validity columns beside it. Reading down the table, internal validity ranges widely across
the roster, from a tight match to the human norm down to corpora that miss it by close to an order of
magnitude, while trait--state direction agreement and rank correlation stay high and roughly flat
throughout. A model's rank on one axis predicts little about its rank on the other axis.
\begin{table}[t]
\centering\footnotesize
\caption{Model leaderboard across both validity criteria, ranked by internal validity (within-person variance error, lower is closer to the human norm). External validity (trait--state direction agreement and rank correlation $\rho$) stays high and roughly flat as internal validity varies, the central split of the benchmark.}
\label{app:leaderboard}
\begin{tabular}{l r r c r r}
\toprule
Model & Params (B) & Var.\ error $\downarrow$ & Levels & Dir.\ (\%) & $\rho$ \\
\midrule
Gemma-2-27B & 27 & 0.043 & 2/6 & 80 & 0.81 \\
GPT-5-5 & -- & 0.061 & 2/6 & 80 & 0.69 \\
Gemma-3-27B & 27 & 0.065 & 3/6 & 90 & 0.84 \\
Qwen-2-5-Coder-32B & 32 & 0.090 & 2/6 & 80 & 0.65 \\
Claude-Sonnet-4 & -- & 0.115 & 2/6 & 80 & 0.63 \\
Qwen3-5-27B & -- & 0.135 & 3/6 & -- & -- \\
GPT-4-1 & -- & 0.140 & 2/6 & 85 & 0.63 \\
Llama-3-1-8B & 8 & 0.145 & 2/6 & 85 & 0.65 \\
Qwen-2-5-72B & 72 & 0.146 & 2/6 & 90 & 0.80 \\
Gemma-3-12B & 12 & 0.147 & 3/6 & 90 & 0.70 \\
Mistral-Small-3-2-24B & 24 & 0.150 & 3/6 & 75 & 0.66 \\
Qwen3-14B & 14 & 0.160 & 2/6 & 80 & 0.68 \\
Deepseek-V3-2 & 671 & 0.161 & 3/6 & 85 & 0.79 \\
Qwen3-30B-A3B-Thinking & 30 & 0.175 & 2/6 & 90 & 0.64 \\
Deepseek-R1-0528 & 671 & 0.186 & 3/6 & 85 & 0.71 \\
Gemini-2-5-Flash & -- & 0.196 & 2/6 & 85 & 0.78 \\
Qwen3-8B & 8 & 0.206 & 2/6 & 90 & 0.57 \\
Deepseek-R1-Distill-Llama-70B & 70 & 0.210 & 2/6 & 90 & 0.74 \\
Llama-4-Maverick & 400 & 0.216 & 2/6 & 80 & 0.71 \\
Qwen3-30B-A3B & 30 & 0.232 & 2/6 & 85 & 0.68 \\
Qwen3-32B & 32 & 0.236 & 2/6 & 85 & 0.72 \\
Llama-3-2-3B & 3 & 0.240 & 2/6 & 80 & 0.61 \\
Llama-3-8B & 8 & 0.242 & 2/6 & 85 & 0.58 \\
Gemini-2-5-Pro & -- & 0.247 & 2/6 & 85 & 0.75 \\
Qwen3-235B-A22B & 235 & 0.250 & 1/6 & 90 & 0.68 \\
GPT-Oss-20B & 21 & 0.270 & 1/6 & 75 & 0.78 \\
GPT-Oss-120B & 117 & 0.287 & 1/6 & 85 & 0.77 \\
Llama-3-2-1B & 1 & 0.309 & 3/6 & 75 & 0.50 \\
Granite-4-1-8B & 8 & 0.315 & 2/6 & 90 & 0.72 \\
Qwen-2-5-7B & 7 & 0.325 & 1/6 & 85 & 0.71 \\
Llama-3-1-70B & 70 & 0.354 & 2/6 & 80 & 0.60 \\
Phi-4 & 14 & 0.362 & 1/6 & 85 & 0.59 \\
Llama-3-3-70B & 70 & 0.363 & 1/6 & 80 & 0.55 \\
Gemma-3-4B & 4 & 0.427 & 3/6 & 100 & 0.83 \\
\bottomrule
\end{tabular}
\end{table}

\end{document}